\documentclass[12pt,letterpaper]{article}
\usepackage{color}
\usepackage{amsmath}
\usepackage{amsthm}
\usepackage{amsfonts}
\usepackage{amssymb}
\usepackage{graphicx}
\usepackage{float}
\usepackage{geometry}
\usepackage[utf8]{inputenc}

\usepackage[all]{xy}
\usepackage{amscd}
\usepackage{amsthm}
\usepackage{longtable}
\usepackage{mathrsfs}
\usepackage{amsmath}
\usepackage{amssymb}
\usepackage{enumitem}
\usepackage[table,dvipsnames]{xcolor}

\usepackage{caption}

\usepackage[utf8]{inputenc}
\usepackage[english]{babel}

\usepackage[thinlines]{easytable}

\usepackage{setspace}
\usepackage{tikz-cd}
\usepackage{tikz}
\usepackage{relsize}

\usepackage{pdfpages}

\usepackage{listings} 
\usepackage{xcolor}

\usepackage{natbib}
\usepackage{comment}
\usepackage{color}
\usepackage[backref=page]{hyperref}
\hypersetup{
 colorlinks=true,
 linkcolor=blue,
 filecolor=magenta,
 citecolor=darkRed,
 urlcolor=cyan,
 pdftitle={Sensitivity analysis for transportability in multi-study, multi-outcome settings},
 }

\usepackage[normalem]{ulem}

\definecolor{codegreen}{rgb}{0,0.6,0}
\definecolor{codegray}{rgb}{0.5,0.5,0.5}
\definecolor{codepurple}{rgb}{0.58,0,0.82}
\definecolor{backcolour}{rgb}{0.95,0.95,0.92}
\definecolor{darkRed}{rgb}{0.60,.03,.03}

\lstdefinestyle{mystyle}{
    backgroundcolor=\color{backcolour},   
    commentstyle=\color{codegreen},
    keywordstyle=\color{magenta},
    numberstyle=\tiny\color{codegray},
    stringstyle=\color{codepurple},
    basicstyle=\ttfamily\footnotesize,
    breakatwhitespace=false,         
    breaklines=true,                 
    captionpos=b,                    
    keepspaces=true,                 
    numbersep=5pt,                  
    showspaces=false,                
    showstringspaces=false,
    showtabs=false,                  
    tabsize=2
}

\theoremstyle{definition}

\newtheorem*{counter example}{Counter Example}

\newtheorem{assumption}{Assumption}

\DeclareUnicodeCharacter{2008}{\textcolor{red}{BAD!!}}

\begin{document}

\def\spacingset#1{\renewcommand{\baselinestretch}%
{#1}\small\normalsize} \spacingset{1}

\title{ {\Large Sensitivity analysis for transportability in multi-study,\\multi-outcome settings}}
\author{ {\normalsize Ngoc Q. Duong, Amy J. Pitts, Soohyun Kim, Caleb H.~Miles }  }
\date{}

\clearpage \maketitle
\thispagestyle{empty}


\begin{center}
    Department of Biostatistics, Mailman School of Public Health, Columbia University 
\end{center}

\begin{abstract}

Existing work in data fusion has covered identification of causal estimands when integrating data from heterogeneous sources. These results typically require additional assumptions to make valid estimation and inference. However, there is little literature on transporting and generalizing causal effects in multiple-outcome setting, where the primary outcome is systematically missing on the study level but for which other outcome variables may serve as proxies. We review an identification result developed in ongoing work that utilizes information from these proxies to obtain more efficient estimators and the corresponding key identification assumption. We then introduce methods for assessing the sensitivity of this approach to the identification assumption.

\end{abstract}	
\noindent%
  \vfill

\doublespacing

\newpage 
\setcounter{page}{1}

\section{Introduction}
\label{section:intro}

Research in clinical medicine and public health is often concerned with estimating the effect of some treatment in a specific target population. However, even in a randomized clinical trial, which is considered the gold-standard study design, ensuring external validity remains a challenge. This can be due to a variety of reasons, including non-random sampling, overly stringent exclusion criteria, or an ill-defined target population of interest \citep{Tan_2022, Kennedy_Martin_2015}. Meta-analysis of summary statistics is a commonly used tool to synthesize and generalize findings from published study-level summary statistics, but tends to rely on strong, often implausible assumptions. An alternative approach that allows for more control over the nuances and heterogeneity across studies is to combine individual-level data, when available, from multiple studies, each of which may contain insufficient information to address a given scientific question by itself, but which collectively have the power to do so. There has been a growing body of work on generalizability and transportability methods, which can help address the problem of external validity of the effect estimates from integrating individual level data across studies.

Generalizability concerns the setting where the study population is a subset of the target population of interest while transportability addresses the setting where the study population is partially or completely external to the target population \citep{Degtiar_2023}. Specifically, generalizability typically involves extending the causal effect estimate derived from a study as long as the covariates in the study population and the target population have common support \citep{Gechter_2015, Tipton_2014}. On the other hand, transportability entails extrapolating the effect estimated from a study in which some primary outcome of interest is observed to a population represented by a sample in which the outcome is not measured.

Existing methodologies involve directly transporting some estimated causal effect, e.g., the average treatmemt effect (ATE), from studies where the outcomes are observed to other studies with missing outcomes or across heterogeneous study designs and settings \citep{Bareinboim_2016, Dong_2020, Pearl_2014, Hunermund_2019}, or to some broader target population \citep{Dahabreh_1_2020, Dahabreh_2_2020, Lesko_2017, Westreich_2017}. When considering multiple studies, it is often the case that one will observe different outcomes at follow up. However, existing methods do not take advantage of these other potentially correlated and informative outcome variables measured at follow-up, which could potentially be leveraged to achieve large efficiency gains. Existing outcome proxy-blind methods typically rely on an assumption of homogeneous conditional potential outcome means for valid transportation of estimation from one population to another. Sensitivity analysis strategies have been proposed to study the extent to which the violation of these assumptions will affect the estimations and inferences drawn \citep{Nguyen_2017, Dahabreh_1_2019, Dahabreh_2022}. 

In ongoing work, we have developed a new strategy to more efficiently estimate the ATE from integrated data across multi-outcome studies, with inconsistent availability of the primary outcome of interest at the study level. The proposed methodology takes advantage of the availability of follow-up measurements of potential correlates of the main outcome to yield more precise estimate of the causal effects. In this article, we consider the key common outcome regression (or conditional exchangeability for study selection) assumption for transportability while leveraging these outcome proxies, which differs slightly from the common outcome regression assumption that has been traditionally used for transportability. We discuss the resulting bias when this assumption is not met, and develop methodology for sensitivity analysis to the violation of this assumption.

The remainder of the article is organized as follows. In Section \ref{section:identification}, we discuss identification of the average treatment effect in the multi-study, multi-outcome setting. In Section \ref{section:bias}, we discuss the bias incurred by violations of the key conditional exchangeability assumption. In Section \ref{section:comparison}, we compare the conditional exchangeability assumption in our setting with that used in settings that do not leverage outcome proxies. In Section \ref{section:sensitivity}, we develop methods for sensitivity analysis for when our assumption is violated. We demonstrate the empirical performance of our proposed methods in a simulation study in Section \ref{section:sims}, and conclude with a discussion in Section \ref{section:discussion}.

\section{Data integration for studies with primary outcome missing systematically}
\label{section:identification}

\subsection{Study and data setting}

In this setting, we let $A$ be the treatment indicator, $W$ be a set of covariates that are commonly observed across studies, $Y$ be the primary outcome variable, the set $\{T_1, \dots, T_k\}$ be all the potential outcome proxies measured at follow-up in any study, and $J_s$ be the study-specific subset of $\{T_1, \dots, T_k\}$ that is measured in study $s$. Suppose there are $\mathcal{S}$ studies that are ordered such that for each $s$ in the first $s^*$ studies, we observe the set of variables $(Y, A, J_s, W)$, while for each $s$ in the remaining $\mathcal{S} - s^*$ studies, only the subset $(A, J_s, W)$ are observed. In other words, $Y$ is systematically missing in the latter set of studies. Unlike the standard setup in other works concerning effect transportability that only involves $(Y, A, W)$, we introduced the use of $\mathcal{T}_s$, where $\mathcal{T}_s \subset J_s$ is some user-specified subset of $J_s$ for each study $s$.  $\mathcal{T}_s$ could be chosen based on availability and subject matter knowledge and must be chosen such that they are observed in at least one of the studies $\{1,2, \dots, s^*\}$. 

Studies can be randomized experiments or observational; however, we will not consider scenarios in which some studies are randomized experiments and others are observational in this work. 
Then the study-specific average treatment effect and conditional average treatment effect can be written as:
\begin{align*}
    ATE(s) &= E(Y_1 - Y_0 \mid S = s) \\
    CATE(w,s) &= E(Y_1 -Y_0 \mid W=w, S = s).
\end{align*}
Accordingly, we can define the overall average treatment effect and conditional average treatment effect as:
\begin{align*}
    ATE &= \sum^{\mathcal{S}}_{s=1} ATE(s) \\
    CATE(w) &= E(Y_1 -Y_0 \mid W = w)
\end{align*}
where the weights can be user-specified such that $\sum_s \pi_s = 1$. For instance, one can choose $\pi_s = P(S = s)$, or the marginal probability of being in each study. Alternatively, we could define $ATE = E_{Q_{W,S}}CATE(W,S)$ for a user-specified, known distribution $Q_{W,S}$ of $W$ and $S$. 

Since $Y$ is not measured in $s \in \{ s^* +1,\dots ,{\mathcal{S}}\}$, we cannot directly estimate the ATE and CATE using data from these studies alone. Our purpose is to transport the ATE from the first $s^*$ studies where $Y$ is observed, to the remaining ${\mathcal{S}} - s^*$ studies while also leveraging the information from the outcome proxy set $\mathcal{T}_s$ to improve efficiency. For ease of notation, let $\sigma_s$ be a subset of the first $s^*$ studies in which both Y and $\mathcal{T}_s$ are observed. We can then use this information from the studies that form $\sigma_s$ to estimate the outcome regression that will allow us to transport the causal effects to study $s$. In this setting, we have shown in ongoing, not-yet-published work that the ATE can be nonparametrically identified as:
\begin{align*} \label{eq:1}
    \Psi^{ATE} &= \sum^{s^*}_{s=1} \pi_s E\{E(Y\mid W, A=1, S = s) - E(Y\mid W,A=0, S = s) \mid S = s \} \\ \tag{1}
    &  + \sum^{{\mathcal{S}}}_{s=s^*+1} \pi_s E[E\{E(Y\mid \mathcal{T}_s, W, A=1, S\in \sigma_s)\mid W, A=1, S = s \} \\
    & \quad \quad \quad \quad \quad - E\{E(Y\mid \mathcal{T}_s, W, A=0, S\in \sigma_s)\mid W, A=0, S = s \} \mid S = s ].
\end{align*}
The terms in the first sum are simply the standard identification formula for the (study-specific) average treatment effects when $Y$ is observed. The second sum is identified since it only depends on the distribution of $Y$ in the studies in $\sigma_s$, i.e., in which $Y$ is actually observed.

Here, we introduced a modification to how transportability has traditionally been done by incorporating information from a set of outcomes measured at follow-up that are correlated with the main outcome of interest. 

\subsection{Assumptions for Identification of the ATE }
This derivation ATE can be nonparametrically identified given the assumptions that are standard for identification for ATE when outcomes are all observed:
\begin{assumption}[Positivity]
\label{assn:positivity}
    $P(A = 1\mid W = w) > 0$ for all $w$ with positive probability.
\end{assumption}
\begin{assumption}[Consistency]
\label{assn:consistency}
    $Y = AY_1 + (1-A)Y_0$.
\end{assumption}
\begin{assumption}[Within-study conditional exchangeability]
\label{assn:exchangeability}
    \[E[Y^a \mid W, A, S=s] = E[Y^a \mid W, S=s] \textrm{ for all } s.\]
\end{assumption}
The validity of our estimator relies on a fourth assumption that allows for the transportation of the effect across studies:
\begin{assumption}[Common outcome regression (proxy-aware version)]
\label{assn:commonOR}
    \[E(Y\mid \mathcal{T}_s, W,A=a,S = s) = E(Y \mid \mathcal{T}_s,W,A=a,S \in \sigma_s) \textrm{ for all } s.\]
\end{assumption}
This is a missing at random (MAR)-type assumption, where $S$ can in a sense be thought of as a missingness indicator, since missingness is systematic by study.

We can also introduce a fifth assumption that is not necessary for identification, but allows for more borrowing of information across studies, which can help with efficiency:
\begin{assumption}[Common distribution of outcome proxies]
\label{assn:commonTdist}
    $\mathcal{T}_s \perp S \mid W, A$ for all $s$.
\end{assumption}
This implies the distribution of $\mathcal{T}_s$ conditional on treatment assignment and baseline covariates is the same across studies. Under this additional assumption, the identification result simplifies to:
\begin{align*}
    ATE &= \sum^{\mathcal{S}}_{s=1} \pi_s E[ E\{ E(Y\mid \mathcal{T}_s, W, A=1, S\in \sigma_s) \mid W, A =1 \} \\
    & \quad - E\{ E(Y\mid \mathcal{T}_s, W, A=0, S\in \sigma_s) \mid W, A =0 \} \mid S = s].
\end{align*}
In ongoing work, we have developed a simple substitution estimator that involves replacing each expectation with a regression-based estimate and the outer expectation with an empirical mean.

For the outcome proxy-blind approach, in addition to the first three standard internal validity assumptions, Assumption \ref{assn:commonOR} is replaced by a slightly different mean outcome exchangeability assumption: across studies assumption (exchangeability over $S$) \citep{Dahabreh_1_2019, Lesko_2017}:
\begin{assumption}[Common outcome regression (proxy-blind version)]
\label{assn:commonORblind}
\begin{align*}
  E(Y\mid W,A=a,S = s) = E(Y \mid W,A=a,S \in \sigma_s) \textrm{ for all } s.
\end{align*}
\end{assumption}
Assumption \ref{assn:commonOR} differs from Assumption \ref{assn:commonORblind} by additionally conditioning on $\mathcal{T}_s$ for each study $s$. Assumptions \ref{assn:commonOR} and \ref{assn:commonTdist} together imply Assumption \ref{assn:commonORblind}. In this article, we will only consider sensitivity analysis for the violation of Assumption \ref{assn:commonOR}. When Assumption \ref{assn:commonTdist} is violated, the ATE estimator based on Assumption \ref{assn:commonOR} (i.e., the substitution estimator based on the identification formula (\ref{eq:1})) will remain consistent.

\section{Characterizing the bias resulting from violation of the identification assumption}
\label{section:bias}
The validity of $\Psi^{ATE}$ is dependent on the key assumption \ref{assn:commonOR}. This assumption requires no heterogeneity in the conditional outcome means given treatment, covariates, and outcomes proxies between studies with and without missing outcome ($Y$) data. This allows for transportation of the conditional outcome means, and correspondingly, the ATE and CATE, estimable from one study to others.

In practice, this could be a strong assumption to make while also untestable using observed data. For instance, in previous unpublished work, we estimated the average treatment effect of cognitive remediation (CR) therapy on Social Behavioral Scale (SBS) score, a measure for social functioning, using harmonized data from three trials in the NIMH Database of Cognitive Training and Remediation Studies (DoCTRS) database. However, the degree of effectiveness of CR, especially on functional and occupational outcomes, was less evident and has been suggested to vary depending on the setting in which the treatment was administered \citep{Barlati_2013, Combs_2008, McGurk_2007, Wykes_2007, Wykes_2011}. When this assumption is violated, the substitution estimators described in the previous section will be biased. Therefore, we examine two strategies for sensitivity analysis in order to examine the robustness of estimates under varying degrees of assumption violation. 

To quantify the degree of violation, let the bias functions be defined as:
\begin{align*} \label{eq:2}
    u(A=1, \mathcal{T}_s, W) &= E(Y\mid \mathcal{T}_s,W,A=1,S = s)-E(Y \mid \mathcal{T}_s, W, A=1, S \in \sigma_s), \\
    u(A=0, \mathcal{T}_s, W) &= E(Y\mid \mathcal{T}_s,W,A=0,S = s)-E(Y \mid \mathcal{T}_s, W, A=0, S \in \sigma_s) \tag{2}
\end{align*}
Then, equation (\ref{eq:1}) when assumption \ref{assn:commonOR} is violated instead becomes:
\begin{align*}
\mathrm{ATE} &=\sum_{s=1}^{s^*} \pi_s E\{E(Y \mid W, A=1, S = s)-E(Y \mid W, A=0, S = s) \mid S = s) \\
&\quad \quad +\sum_{s=s^*+1}^{\mathcal{S}} \pi_s E\left[E\left\{E\left(Y \mid \mathcal{T}_s, W, A=1, S \in \sigma_s\right) \mid W, A=1, S = s\right)\right\} \\
&\quad \quad \quad \quad \left.\left.-E\left\{E\left(Y \mid \mathcal{T}_s, W, A=0, S \in \sigma_s\right) \mid W, A=0, S = s\right)\right\} \mid S = s\right] \\
&\quad \quad +\sum_{s=s^*+1}^{\mathcal{S}} \pi_s E[E\left\{u\left(A=1, \mathcal{T}_s, W\right) \mid W, A=1, S = s\right\} \\ 
&\quad \quad \quad \quad -E\left\{u\left(A=0, \mathcal{T}_S, W\right) \mid W, A=0, S = s\right\} \mid S = s],
\end{align*}
where the last sum is not identified. Then, the study-specific bias for study $s$ is:
\begin{align*} \label{eq:3}
&E\left[E\left\{u\left(A=1, \mathcal{T}_s, W\right) \mid W, A=1, S = s\right\}- E\left\{u\left(A=0, \mathcal{T}_s, W\right) \mid W, A=0, S = s\right\}  \mid S=s \right]  \\
=& \; E[\delta^*(W) |S=s]. \tag{3}
\end{align*}

By rearranging terms, $\delta^*(W)$ can be alternatively written as: 
\begin{align*} \label{eq:4}
&E\left[E\left(Y \mid \mathcal{T}_s, W, A=1, S = s\right) - E\left(Y \mid \mathcal{T}_S, W, A=1, s \in \sigma_s\right) \mid W, A=1, S = s\right] \\
& \quad \quad - E\left[E\left(Y \mid \mathcal{T}_s, W, A=0, S = s\right)-E\left(Y \mid \mathcal{T}_s, W, A=0, s \in \sigma_s\right) \mid W, A=0, S = s\right] \\
& =E(Y \mid W, A=1, S = s)-E(Y \mid W, A=0, S = s) \\
&\quad \quad -\{E\left[E\left(Y \mid \mathcal{T}_s, W, A=1, s \in \sigma_s\right) \mid W, A=1, S = s\right] \\ 
&\quad \quad\quad \quad- E\left[E\left(Y \mid \mathcal{T}_s, W, A=0, s \in \sigma_s\right) \mid W, A=0, S = s\right]\}. \tag{4}
\end{align*}
The latter term cannot be simplified unless Assumption \ref{assn:commonTdist} holds.

\section{Comparison with bias functions in settings without incorporation of follow-up surrogate outcomes}
\label{section:comparison}
In recent work, \cite{Dahabreh_1_2019} developed sensitivity analysis for transportability considering a similar setting of two types of studies with and without missing outcomes. In the base case, there are two studies considered (missingness of the outcome variable denoted by a binary indicator $S$). To describe this setting using our notation,  we simply have $\sigma_0 = \sigma_1 = \{1\}$ (i.e., study $S = 1$ with the observed outcome of interest is used to impute the conditional outcome means for study $S = 0$). Equivalently, for ease of interpretation in the base case, let $S = 1$ and $S = 0$ denote the study where the primary outcome of interest is observed and not observed, respectively.

In the setting where the model used to impute conditional potential outcomes does not utilize information from $\mathcal{T}_s$, 
\cite{Dahabreh_1_2019} define:
\begin{equation*}
    u(A=a, W) = E[Y \mid A=a, W, S = 1] - E[Y \mid A=a, W,  S = 0].
\end{equation*}
The difference between these bias functions can then be obtained as:
\begin{align*}
    \delta(W) &= u(A=1, W) - u(A=0,W) \\
    &= E[Y^1-Y^0 \mid W, S = 1] - E[Y^1-Y^0 \mid W, S = 0]
\end{align*}

This expression can be qualitatively expressed as the difference in the conditional average treatment effects between the two studies. This qualitative interpretation can aid in conceptualizing and thinking about more appropriate values and range for sensitivity parameters when examining robustness of the results. More specifically, assuming higher levels of the outcome are preferred, if we believe the participants in studies with missing outcomes benefit less from treatment, then true $\delta$ can be assumed to be positive and vice versa \citep{Dahabreh_1_2019}. Since our bias functions are conditional on the set of proxy outcomes, the term $\delta^*(W)$ in (\ref{eq:4}) unfortunately cannot be reduced further to a more interpretable statistical entity. When we take $\mathcal{T}_s$ to be the empty set, the bias function $\delta^*(W)$ reduces to the same expression. 

\section{Accounting for violation of the common outcome regression assumption through sensitivity analyses}
\label{section:sensitivity}
We consider two scenarios in which we assume the bias terms $u(A=1, \mathcal{T}_s, W)$ and $u(A=0, \mathcal{T}_s, W)$ to be 1) constants and 2) bounded functions of the outcome proxies and/or baseline covariates. The first scenario involves making a stronger assumption about the bias terms. On the other hand, the second scenario requires weaker assumptions but allow them to be non-constant.

\subsection{Bias functions assumed to be some fixed values}
Although it might be more reasonable to assume that the bias functions are dependent on some baseline covariates, for ease of implementation of sensitivity analysis, one can also suppose they are constant. When $u(A=1, \mathcal{T}_s, W)$ and $u(A=0, \mathcal{T}_s, W)$ are independent of the baseline covariates $W$ and the outcome proxy set $\mathcal{T}_s$, the conditional expectations of the bias functions, and in turn, the term $\delta^*(W)$ in (\ref{eq:3}), reduce to:
\begin{equation*} \label{eq:5} \tag{5}
    \delta = u_1 - u_0, \textrm{ where } \delta,   u_1, \textrm{ and } u_0 \in \mathbb{R}
\end{equation*}
The sensitivity analysis involves correcting for the above-mentioned bias term by adding it back to the identification formula $\Psi^{ATE}$, which relies on the common outcome regression assumption.   

\begin{align*} 
ATE =&\sum_{s=1}^{{s^*}} \pi_s E\left\{E\left(Y \mid W, A=1, S \in \sigma_s\right)-E\left(Y \mid W, A=0, S \in \sigma_s\right) \mid S = s\right\} \\
&+\sum_{s=s^*+1}^{{{\mathcal{S}}}} \pi_s E\left[E\left\{E\left(Y \mid \mathcal{T}_s, W, A=1, S \in \sigma_s\right) \mid W, A=1, S = s\right\}\right. \\
&-\left.E\left\{E\left(Y \mid \mathcal{T}_s, W, A=0, S \in \sigma_s\right) \mid W, A=0, S = s\right\} \mid S = s\right]+\sum_{s=s^*+1}^{{{\mathcal{S}}}} \pi_s\left(u_1-u_0\right) \\ \label{eq:6} \tag{6} 
=& \Psi^{ATE}+\sum_{s=s^*+1}^{{\mathcal{S}}} \pi_s\left(u_1-u_0\right)
\end{align*}
where $u_1$ and $u_0$ are scalars.

In practice, the true bias term would be unknown. Thus, one strategy is to propose a grid of sensitivity parameters that covers the potential range of values in which the true bias term might fall. This grid of sensitivity parameters can be specified using subject-matter knowledge. We can then adjust for the bias term in the estimation step by adding back the different sensitivity parameters to the estimated ATE using our proposed method. This also allows for observation of the behavior of the estimated ATE as we vary the sensitivity parameters.

\subsection{Bounded covariate-dependent bias functions }

One might also believe that the bias term is not constant at all levels of the baseline covariates and/or the outcome proxies. When the assumption of fixed-value bias terms is considered too strong, but the functional forms for bias terms cannot be confidently determined from existing knowledge of the data mechanism (as will typically be the case), one can still recover some information about the true ATE without having to correctly specify the bias terms. If we instead assume the bias terms to be some bounded functions, we can compute a bound around the (naïve) ATE estimate that contains the true ATE by varying the bounds of these functions. This provides information on how far away the true ATE can be from the estimate obtained constrained by the bounds of the bias term.

Identifying the bounds for the bias term can be expressed as maximizing and minimizing the objective function: 
\begin{equation*}
    E[E[u(A=1, \mathcal{T}_s, W) \mid W, A=1, S=s] -E[u(A=0, \mathcal{T}_s, W) \mid W, A=0, S=s] \mid S=s]
\end{equation*}
subject to the following constraints: 
\begin{align*}
    |u(A=1, \mathcal{T}_s = t_s, W = w)| &\leq \gamma_1 \\
    |u(A=0, \mathcal{T}_s = t_s, W = w)| &\leq \gamma_0 
\end{align*} for all $t_s$ and $w$, which implies $| E[u(A=1, \mathcal{T}_s, W) \mid W, A=1, S = s ]| \leq \gamma_1$ and $|E[u(A=0, \mathcal{T}_s, W) \mid W, A=0, S = s ] | \leq \gamma_0$ where $\gamma_1, \gamma_1 \in \mathbb{R}^+$. 

Then we have $-(\gamma_1 + \gamma_0) \leq u(A=1, \mathcal{T}_s, W) - u(A=0, \mathcal{T}_s, W) \leq \gamma_1 + \gamma_0$. If we have no reason to suspect we know more about the bounds of one bias function than the other (as will typically be the case), we may simply choose to specify a scalar sensitivity parameter $\gamma$ to be the maximum of $\gamma_1$ and $\gamma_2$, in which case we have $-2 \gamma \leq u(A=1, \mathcal{T}_s, W) - u(A=0, \mathcal{T}_s, W) \leq 2\gamma $. 

By equation (\ref{eq:6}) even though we do not know the form of the bias functions $u(A=1, \mathcal{T}_s, W)$ and $u(A=0, \mathcal{T}_s, W)$, we can partially recover the true ATE using the bounds around the naïve estimate:
\begin{align*} \label{eq:7}
    \Psi^{ATE} -2 \max (\gamma_1, \gamma_0) &\leq  ATE \leq \Psi^{ATE} + 2 \max (\gamma_1 , \gamma_0) \\
    \Psi^{ATE} -2 \gamma &\leq ATE \leq \Psi^{ATE} + 2 \gamma  \tag{7}
\end{align*}
If the bias functions are in fact bounded by some value smaller than or equal to our specified values for the sensitivity bounds, the true ATE would fall between $[\Psi^{ATE} -2\gamma, \Psi^{ATE} +2\gamma]$. Then, the true ATE is partially identified without assumptions about the functional form of $u(A=1,\mathcal{T}_s,W)$ and $u(A=0,\mathcal{T}_s,W)$. One can then use the bootstrap standard error for the substitution estimator of the identification formula (\ref{eq:1}) to determine the amount to add and subtract from the upper and lower bounds, respectively, in order to produce confidence intervals for the partial identification sets for each value of the sensitivity parameter. Since the sensitivity bounds are a deterministic function of the sensitivity parameter, bootstrapping need only be done once.

\section{Simulations}
\label{section:sims}
\subsection{Data generating mechanism}
We consider the setting of two studies, with $S$ = 1 indicating the study where the primary outcome is available. We generate random sample draws with sample size n = 100 for both studies. The data generating mechanism is as follows. $W, T_0$ come from independent standard normal distributions, and $T_1$ comes from a normal distribution with mean and variance of 1. Then 
\begin{align*}
    T &= I(A=1) \times T_1 + I(A=0) \times T_0 \\
    Y^0 &= -4 T_0 + W + \epsilon_0 \\
    Y^1 &= 4 T_1 + W + \epsilon_1 \\
    Y &= I(A=1) \times Y_1 + I(A=0) \times Y_0
\end{align*}
where $\epsilon_1, \epsilon_0 \sim N(0,1)$. 

Via these specifications, $T$ fully mediates the relationship between $A$ and $Y$ (direct effect from $A$ to $Y$ is constrained to be 0). As a result, the true ATE $= 4$. This is also a more basic setting in which the vector T is observed in all studies.

Due to the nature of the DoCTRS database, which is comprised of randomized clinical trials, in our base setting, we specified the marginal probability P(A = 1) = 0.5, representing random treatment assignment. This treatment assignment satisfies the positivity and exchangeability assumption.

Specifically, to incorporate the difference in conditional outcome means between the two types of studies, in studies missing the outcome, we added constant bias terms to the counterfactual outcomes $Y_0$ and $Y_1$. Similar to the data generating step, we preserved the observed counterfactual outcome from the corresponding treatment assignment, which satisfies the consistency assumption. By (\ref{eq:5}), we have:
\begin{align*} \label{eq:9}
Y_{S = 1}^0&=Y_{S = 0}^0+u_0 \\ \tag{8}
Y_{S = 1}^1&=Y_{S = 0}^1+u_0+\delta
\end{align*}
for $u_0 \in \{-3,0,3\}, \delta \in \{-2,0,2\}$.

Then the bias reduces to a single parameter $\delta$, since it is no longer a function of $u_0$ when computing the ATE:
\begin{equation*} \label{eq:10} \tag{9}
    E(Y^1-Y^0\mid S=1)=E(Y^1-Y^0\mid S=0)+\delta
\end{equation*}
In the case where the bias term is a function of baseline covariates and surrogate outcome, we had the following specification for the true bias:
$$
\begin{array}{l}
u_0=b_0 \times \sin \left(T_s + W\right) \\
u 1=b_1 \times \frac{\exp \left(T_s+W\right)}{1+\exp \left(T_s+W\right)}
\end{array}
$$
for $b_0 \in\{2,3,4\}$ and $b_1 \in\{1,2,3\}$.

\subsection{ Adjusting for sensitivity parameter in estimation step }
In the presence of non-zero bias, when the value of the sensitivity parameter $\delta$ is specified such that it is equal to true $\delta$, the ATE estimate after bias adjustment tends to be closer to the true ATE after compared to before. In addition, the corresponding 95\% CIs are expected to cover the true ATE 95\% of the times. Although coverage probability can be examined more in a more robust fashion using bootstrapped confidence intervals across all simulations, in Fig.~\ref{fig:fig1}, \ref{fig:fig2}, and \ref{fig:fig7}-\ref{fig:fig10}, the 95\% CIs covers the true ATE at the value of the sensitivity parameter that reflects the degree of assumption violation all but one instance, which is in line with our expectations. 

\textbf{Scenario 1.} When the bias terms are assumed to be constants, a natural approach would be to specify a two-dimensional grid of sensitivity parameters for both scalars $u_0$ and $u_1$. However, by (8), it is equivalent to specifying $u_0$ (or $u_1$) and $\delta$. In fact, since the $u_0$ (or $u_1$) as constant terms cancel out during adjustment, it is sufficient to specify one sensitivity parameter $\delta$ (9). We also note that $\delta$ being 0 does not necessarily imply assumption \ref{assn:commonOR} is met, since the bias terms $u_0$ and $u_1$ could cancel exactly.

To implement sensitivity analysis, we follow the steps:
\begin{enumerate}
    \item Specify a grid of sensitivity parameters $\delta$. The grid should be reasonably wide to contain true $\delta$.
    \item Estimate the naïvely transported ATE using the identification result in (\ref{eq:1})
    \item Sequentially add the values in the sensitivity parameter grid to the naïvely estimated ATE, using the result in (\ref{eq:6}) to obtain the bias-corrected ATE estimates.
\end{enumerate}
We then plotted the bias-corrected estimates under different sensitivity parameters against the true ATE. Additionally, we bootstrapped the bias-corrected estimates to obtain the 95\% confidence intervals and explore coverage across different values of $u_0$ and $\delta$.	

\textbf{Scenario 2.} When we want to make minimal assumptions about the functional form of the bias, we can still perform sensitivity analysis on the true ATE using the following steps: 
\begin{enumerate}
    \item Specify a grid of sensitivity parameters called $\gamma$ that potentially include the upper and lower bounds of the true bias functions
    \item Computed the “naïve” ATE estimate using the identification result in (\ref{eq:1})
    \item Construct the upper and lower bound around the estimated ATE using (\ref{eq:7}) where $\gamma$ is replaced with the sensitivity parameters. 
\end{enumerate}
We also plot the naïve ATE estimates and the bounds around these estimates at each value of the sensitivity parameters. In practice, the bias functions are of course unknown and cannot be estimated from observed data. Therefore, when specifying the grid of sensitivity parameters, the analyst needs to employ subject matter knowledge about the data generating mechanism to select values of $\delta$ and $\gamma$. 

We then explore the behavior of the adjusted estimators via simulations. In the first case, we focused on the general unbiasedness of the correctly-adjusted point estimate for both the overall ATE and ATE among studies with missing outcomes, as well as the 95\% CI coverage across degrees of assumption violation (i.e., across values of true $u_0$ and $\delta$). In the second case, we looked for correct bounding of the true ATE.

\subsection{Simulation Results}

\subsubsection{Bias terms as constants}

\begin{figure}[H]
    \centering
    \includegraphics[scale=0.7]{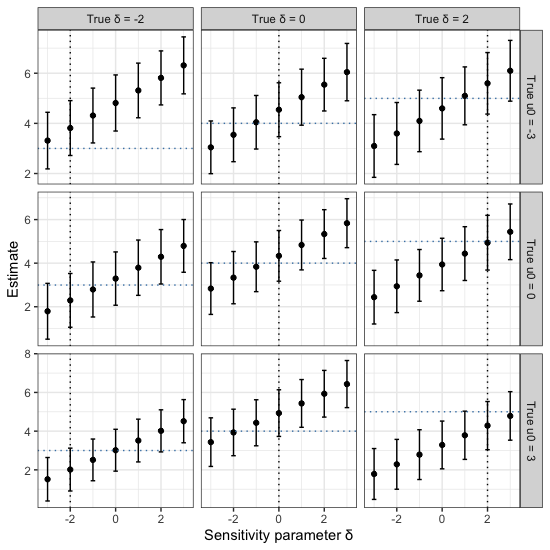}
    \caption{Sensitivity-parameter-adjusted ATE estimate shown against the true overall ATE across values of the true bias and sensitivity parameter; n=100 for each study, 95\% CI constructed from 1000 bootstrap samples. When sensitivity parameter $\delta$ = 0, the adjusted estimate corresponds to the unadjusted estimate. Horizontal dotted line shows the true ATE given true $\delta$; vertical dotted line indicates sensitivity parameter $\delta$ equals true $\delta$}
    \label{fig:fig1}
\end{figure}

\begin{figure}[H]
    \centering
    \includegraphics[scale=0.7]{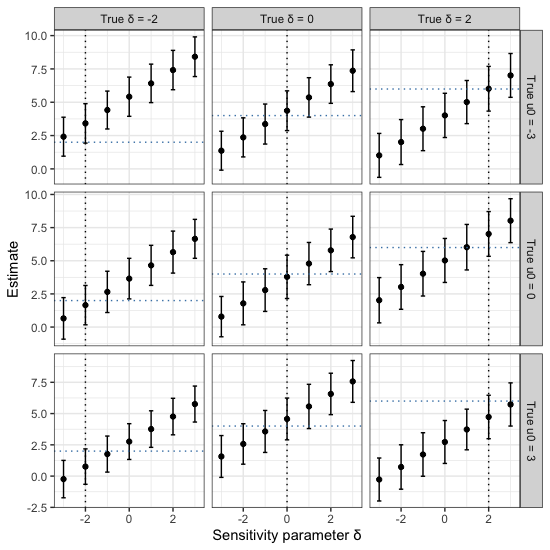}
    \caption{Sensitivity-parameter-adjusted ATE estimates shown against the true study-specific ATE in the study in which the outcome is unobserved across values of the true bias and sensitivity parameter; n=100 for each study, 95\% CI constructed from 1000 bootstrap samples. When sensitivity parameter $\delta$ = 0, the adjusted estimate corresponds to the unadjusted estimate. Horizontal dotted line shows the true study-specific ATE given true $\delta$; vertical dotted line indicates sensitivity parameter $\delta$ equals true $\delta$}
    \label{fig:fig2}
\end{figure}

We examine the estimates produced by our method under the different degrees of violation of assumption \ref{assn:commonOR}, before and after taking into account the specified sensitivity parameter. Figure \ref{fig:fig1} shows the estimates (95\% CI) for the true overall ATE using our method under varying magnitudes and directions of the bias terms from one single simulation. 
In the presence of non-zero bias, when the value of the sensitivity parameter $\delta$ is specified such that it is equal to true $\delta$, the ATE estimate after bias adjustment tends to be closer to the true ATE after compared to before. In addition, the corresponding 95\% CIs are expected to cover the true ATE 95\% of the times. Although coverage probability can be examined more in a more robust fashion using bootstrapped confidence intervals across all simulations, in Fig.~\ref{fig:fig1}, \ref{fig:fig2}, and \ref{fig:fig7}-\ref{fig:fig10}, the 95\% CIs covers the true ATE at the value of the sensitivity parameter that reflects the degree of assumption violation all but one instance, which is in line with our expectations. 

Figure \ref{fig:fig2} shows similar results for the study-specific ATE estimates in the study with missing outcomes (before and after bias adjustment) from the same simulated data. Compared to the results in Figure \ref{fig:fig1}, after adjustment using the correct sensitivity parameters, the 95\% CIs contain the true ATE more frequently than the CIs of the unadjusted estimates in the study with missing primary outcome. Figure 2 also shows an example where inference is sensitive to the violation of our assumption at a magnitude of $\delta$ between -1 and -2 ($u_0 = -3$, bottom left panel), between which the 95\% CI changes from not containing to zero to containing zero. 

When we increased the sample size (n=200 and n=500), we saw general reductions in the errors of these single estimates (Figures \ref{fig:fig7}, \ref{fig:fig9}). In most cases, even when there is error in the adjusted estimates, the 95\% CI bootstrap confidence intervals provide good coverage (Figures \ref{fig:fig1}, \ref{fig:fig7}, \ref{fig:fig9}). The reduction in error and improved coverage are more pronounced when estimating the study-specific effect in the study with missing outcomes than in the overall ATE combining the two studies (Figures \ref{fig:fig8}, \ref{fig:fig10}). 

\begin{figure}[H]
    \centering
    \includegraphics[scale=0.7]{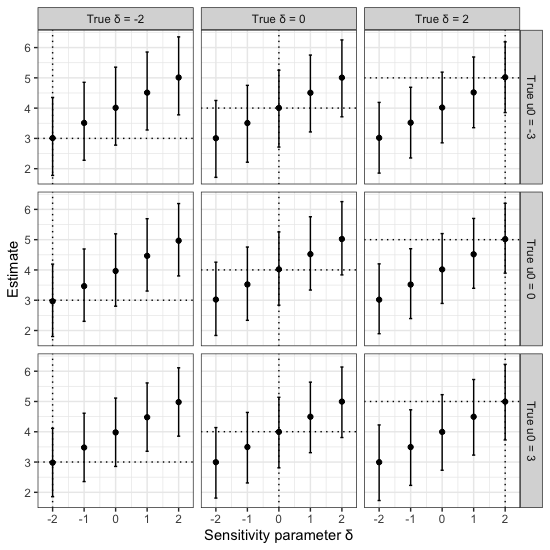}
    \caption{Sensitivity-parameter-adjusted ATE estimates shown against the true overall ATE across values of the true bias sensitivity parameter; mean, 2.5th and 97.5th quantiles obtained from 1000 simulations. When sensitivity parameter $\delta$ = 0, the adjusted estimate corresponds to the unadjusted estimate. Horizontal dotted line shows the true overall ATE given true $\delta$; vertical dotted line indicates sensitivity parameter $\delta$ equals true $\delta$}
    \label{fig:fig3}
\end{figure}

\begin{figure}[H]
    \centering
    \includegraphics[scale=0.7]{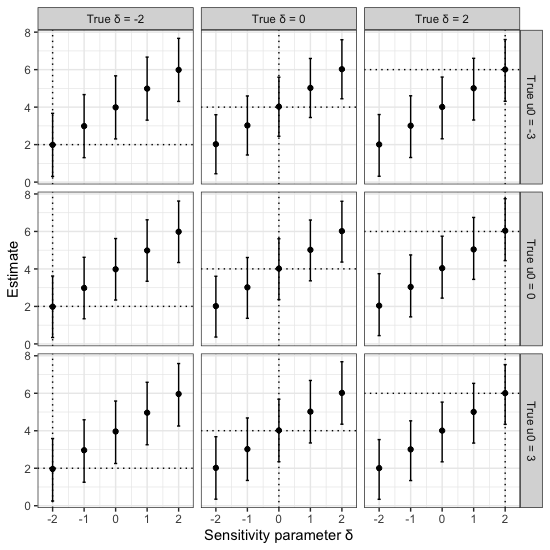}
    \caption{Sensitivity-parameter-adjusted ATE estimates shown against the true ATE in the study with missing outcome across values of the true bias and sensitivity parameter; mean, 2.5th and 97.5th quantiles obtained from 1000 simulations. When sensitivity parameter $\delta$ = 0, the adjusted estimate corresponds to the unadjusted estimate. Horizontal dotted line shows the true study-specific ATE given true $\delta$; vertical dotted line indicates sensitivity parameter $\delta$ equals true $\delta$}
    \label{fig:fig4}
\end{figure}

We also ran 1000 simulations under the same data generating mechanism and obtained the unadjusted and sensitivity-parameter-adjusted estimates for each simulation. We then showed the mean and 2.5th and 97.5th quantiles of these estimates under each combination of the true bias values. We can see that when averaged across 1000 simulations, the adjusted estimates closely approximate the true ATE (Figures \ref{fig:fig3}, \ref{fig:fig4}) when the true value of $\delta$ is used for the sensitivity parameter. 
When approximate sensitivity parameters $\delta$ are used ($\delta \in\{-1,1\}$ when true $\delta \in\{-2,2\}$), the middle 95\% values of adjusted estimates also cover the true ATE whereas those of unadjusted estimates do not (Figure \ref{fig:fig4}).

\begin{figure}[H]
    \centering
    \includegraphics[scale=0.7]{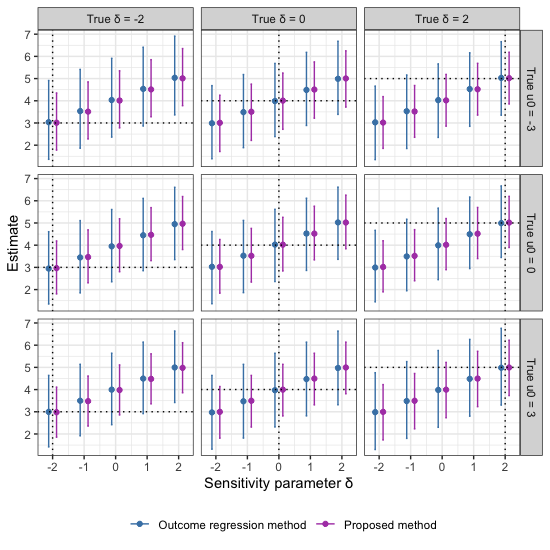}
    \caption{Sensitivity-parameter-adjusted ATE estimates obtained from our proposed method and the outcome proxy-blind method; mean, 2.5th and 97.5th quantiles obtained from 1000 simulations. When sensitivity parameter $\delta$ = 0, the adjusted estimate corresponds to the unadjusted estimate. Horizontal dotted line shows the true overall ATE given true $\delta$; vertical dotted line indicates sensitivity parameter $\delta$ equals true $\delta$}
    \label{fig:fig5}
\end{figure}

Figure \ref{fig:fig5} compares the errors in the estimates and sensitivity of associated inferences between the outcome proxy-blind method of \cite{Dahabreh_2_2020, Lesko_2017} and our proposed method across 1000 simulations. 
The distributions of the estimates from both methods are centered on the true parameter. However, the estimates tend to be more precise when we utilize the information from the outcome proxy (as demonstrated through the narrower 2.5th-97.5th quantile range). The efficiency gains have implications for the sensitivity analysis, since resulting inferences are not as sensitive given analogous magnitude in violation of the identification assumption \ref{assn:commonOR}. 

Assumption \ref{assn:commonOR} implies both $u_0$ and $u_1$ equal 0. As a result, the true $\delta$ also equals 0. This suggests transportation of the conditional potential outcome means, and in turn, the conditional average treatment effects, can be done without incurring bias (vertical middle panes, figure  \ref{fig:fig3}). We also observed that, when $\delta$ is 0, regardless of the values of $u_0$ (and $u_1$), there is also no bias (vertical middle panes, figure \ref{fig:fig3}) in the unadjusted estimator. In both cases, no bias correction would be necessary, and incorporating a non-zero $\delta$ sensitivity parameter will actually introduce bias to the estimate. 

\subsubsection{Bias terms as bounded functions}

\begin{figure}[H]
    \centering
    \includegraphics[scale=0.7]{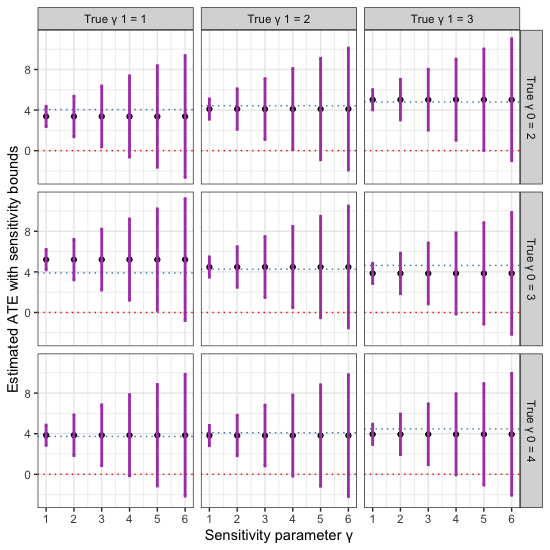}
    \caption{ATE estimates with sensitivity bounds shown against the true overall ATE across values of the true bias and sensitivity parameter. When sensitivity parameter $\gamma$ = 0, the bounds collapse to a point estimate. Blue horizontal dotted line shows the true study-specific ATE given true bias functions }
    \label{fig:fig6}
\end{figure}

When the sensitivity parameter $\gamma$ is greater or equal to $\max\{\gamma_0, \gamma_1\}$ for the true function bounds $\gamma_0$ and $\gamma_1$, the bounds always include the true ATE when the bias functions are bounded by $\gamma_0$ and $\gamma_1$ (Figure  \ref{fig:fig6}). 
Although this approach requires minimal assumptions about the bias functional form, it can also be conservative since the true bias functions are unlikely to evaluate to the bounds across the domain of the functions. For instance, the bottom three panels of Figure \ref{fig:fig6} show that when the sensitivity parameter $\gamma$ is greater than or equal to max(true $\gamma_0$, true $\gamma_1$), while the bounds on the estimate contain the true ATE, they also contains the null value zero as well. On the other hand, these bounds do not rely on an assumption of constant bias functions, which we may often have no reason to believe. Here, we demonstrated through simulations that sensitivity analysis with relaxed and more credible assumptions can still provide helpful information about the parameter of interest. However, when the bounds are too narrow or too wide, sensitivity analysis using bounded bias functions might not be accurate (i.e., not containing the true parameter) or useful (i.e., containing the null value when the truth is non-null), respectively.

\section{Discussion }
\label{section:discussion}
In this paper, we discussed a data integrative method that utilizes information from available proxies of the outcome of interest measured at follow-up for efficiency gains. We then presented two sensitivity analysis strategies specific to this approach for causal effect transportation when the identification assumption is violated. Our modification to the identification of the ATE in (\ref{eq:1}) allows for more efficient estimators given sufficiently strong outcome proxies. As a result, our bias functions also have similar, yet distinct interpretations than the bias functions of \cite{Dahabreh_1_2019}. 

When the bias terms are assumed to be constants, we can obtain different bias-adjusted point estimates based on our specification of the sensitivity parameters. Additionally, via obtaining the 95\% bootstrap confidence interval for the bias-adjusted estimates, we can examine the robustness of inferences made using our method under varying magnitudes of assumption violation. Specifically, beyond certain values of the sensitivity parameters, the 95\% CI will cross the null value 0. These are the degrees of violation that can affect inferences (where the 95\% CI suggest a change from significant results to non-significant results).

We also proposed sensitivity analysis using bounded bias functions as an alternative when one believes the assumption of a fixed-value bias term is too strong. This approach allows for inferences with minimal assumptions about the unobserved bias functions but can still provide useful information about the parameter of interest. Due to fewer assumptions being made, the results are more conservative and robust, hence more reasonable and credible. Specifically, although we are unable to obtain a point estimate, sensitivity analysis using bounded bias functions can still be informative in the sense of providing information about the general direction of the parameter of interest (beneficial or harmful). This method is generally more conservative if the bounds on the functions are not close to their extreme values, if the bias functions are generally not close to their extreme values, or if there is a large difference between the extrema of the two bias functions.

Correct specification of the bias functions would allow for more precise and informative estimation of the true ATE. However, since they are generally unknown and non-estimable from observed data, sensitivity analysis will typically be the realistic course of action.

When conducting sensitivity analysis, the analyst can start off by specifying a wide grid of the sensitivity parameter and examining the behaviors of the point estimates and 95\% CI (first approach) as well as bounds around the estimates (second approach). They can then search for the “critical” sensitivity parameters that still suggest rejection of the null hypothesis, i.e., the 95\% CI (in the first case) and bounds around the estimate (in the second case) that do not contain 0. It can be determined if greater bias is plausible by using background knowledge of the data generating mechanism or further hypothesizing about such mechanism. If there is little or no evidence that the true bias functions exceed these critical sensitivity parameters, one can be more comfortable in concluding that the observed effect and associated inferences are robust to violation of the transportability assumption \citep{Ding_2016, Cornfield_1959}.

\bibliographystyle{apalike}
\bibliography{references}

\newpage 
\appendix
\renewcommand\thefigure{\thesection.\arabic{figure}}     
\section{Additional figures}
\setcounter{figure}{0}

\begin{figure}[H]
    \centering
    \includegraphics[scale=0.7]{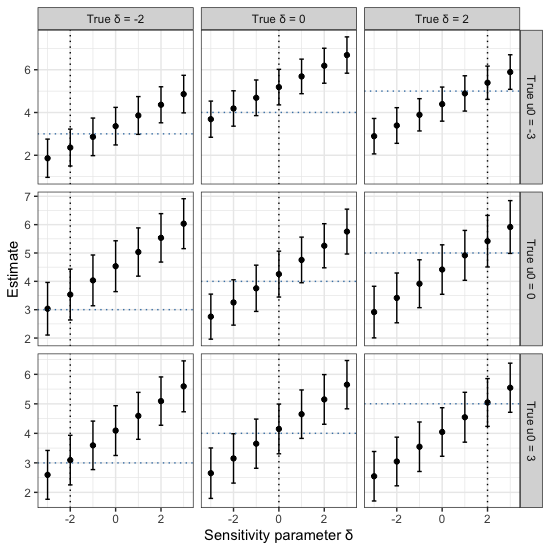}
    \caption{Sensitivity-parameter-adjusted ATE estimates shown against the true overall ATE across values of the true bias and sensitivity parameter; n=200 for each study, 95\% CI constructed from 1000 bootstrap samples. When sensitivity parameter $\delta$ = 0, the adjusted estimate corresponds to the unadjusted estimate. Horizontal dotted line shows the true overall ATE given true $\delta$; vertical dotted line indicates sensitivity parameter $\delta$ equals true $\delta$}
    \label{fig:fig7}
\end{figure}

\begin{figure}[H]
    \centering
    \includegraphics[scale=0.7]{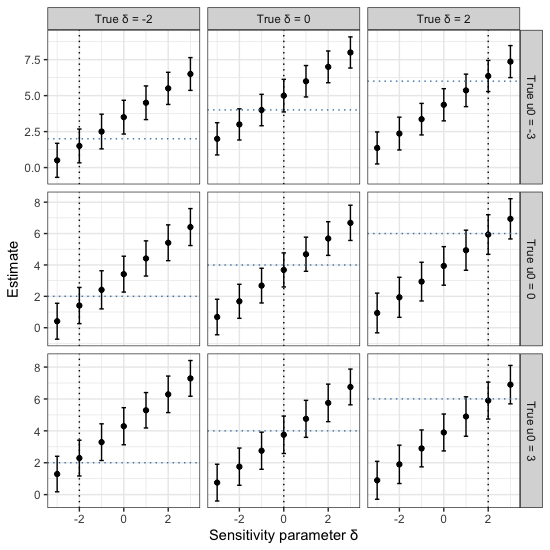}
    \caption{Sensitivity-parameter-adjusted ATE estimates shown against the true ATE in the study with missing outcome across values of the true bias and sensitivity parameter; n=200 for each study, 95\% CI constructed from 1000 bootstrap samples. When sensitivity parameter $\delta$ = 0, the adjusted estimate corresponds to the unadjusted estimate. Horizontal dotted line shows the true study-specific ATE given true $\delta$; vertical dotted line indicates sensitivity parameter $\delta$ equals true $\delta$ }
    \label{fig:fig8}
\end{figure}

\begin{figure}[H]
    \centering
    \includegraphics[scale=0.7]{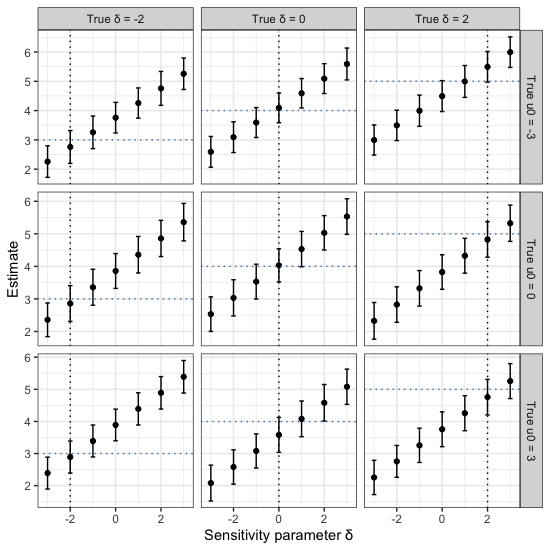}
    \caption{Sensitivity-parameter-adjusted ATE estimates shown against the true overall ATE across values of the true bias and sensitivity parameter; n=500 for each study, 95\% CI constructed from 1000 bootstrap samples. When sensitivity parameter $\delta$ = 0, the adjusted estimate corresponds to the unadjusted estimate. Horizontal dotted line shows the true overall ATE given true $\delta$; vertical dotted line indicates sensitivity parameter $\delta$ equals true $\delta$}
    \label{fig:fig9}
\end{figure}

\begin{figure}[H]
    \centering
    \includegraphics[scale=0.7]{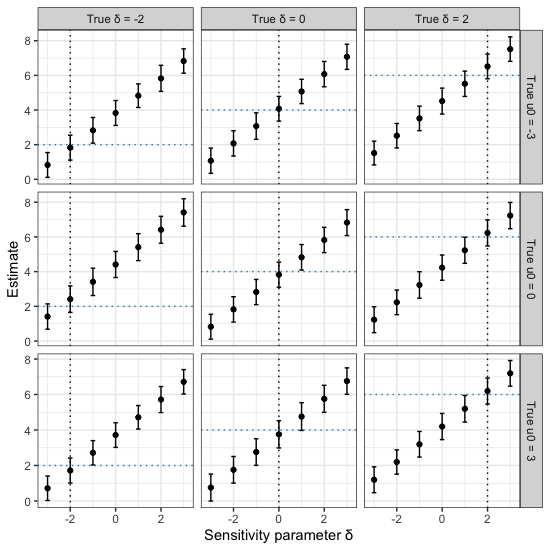}
    \caption{Sensitivity-parameter-adjusted ATE estimates shown against the true ATE in the study with missing outcome across values of the true bias and sensitivity parameter; n=500 for each study, 95\% CI constructed from 1000 bootstrap samples. When sensitivity parameter $\delta$ = 0, the adjusted estimate corresponds to the unadjusted estimate. Horizontal dotted line shows the true study-specific ATE given true $\delta$; vertical dotted line indicates sensitivity parameter $\delta$ equals true $\delta$}
    \label{fig:fig10}
\end{figure}

\newpage 
\section{Derivation of the sensitivity analysis formula when Assumption \ref{assn:commonOR} is violated}

\begin{align*}
\mathrm{ATE}&=\sum_{s=1}^{\mathrm{s^*}} \pi_s E\{E(Y \mid W, A=1, S = s)-E(Y \mid W, A=0, S = s) \mid S = s\} \\
&\quad \quad + \sum_{s=s^*+1}^{S} \pi_s E\left[E\left\{E\left(Y \mid \mathcal{T}_s, W, A=1, S = s\right) \mid W, A=1, S = s\right\}\right. \\
&\quad \quad \quad \quad - \left.E\left\{E\left(Y \mid \mathcal{T}_s, W, A=0, S = s\right) \mid W, A=0, S = s\right\} \mid S = s\right] \\
&=\sum_{s=1}^{s^*} \pi_s E\{E(Y \mid W, A=1, S = s)-E(Y \mid W, A=0, S = s) \mid S = s\} \\
& \quad \quad +\sum_{s=s^*+1}^{{\mathcal{S}}} \pi_s E\left[E\left\{E\left(Y \mid \mathcal{T}_s, W, A=1, S \in \sigma_s\right)+u\left(A=1, \mathcal{T}_s, W\right) \mid W, A=1, S = s \right\}\right. \\
&\quad \quad \quad \quad - \left.E\left\{E\left(Y \mid \mathcal{T}_s, W, A=0, S \in \sigma_s\right)+u\left(A=0, \mathcal{T}_s, W\right) \mid W, A=0, S = s\right\} \mid S = s\right] \\
&=\sum_{s=1}^{s^*} \pi_s E\{E(Y \mid W, A=1, S = s)-E(Y \mid W, A=0, S = s) \mid S = s) \\
&\quad \quad +\sum_{s=s^*+1}^{{\mathcal{S}}} \pi_s E\left[E\left\{E\left(Y \mid \mathcal{T}_s, W, A=1, S \in \sigma_s\right) \mid W, A=1, S = s\right)\right\} \\
&\quad \quad \quad \quad \left.\left.-E\left\{E\left(Y \mid \mathcal{T}_s, W, A=0, S \in \sigma_s\right) \mid W, A=0, S = s\right)\right\} \mid S = s\right] \\
&\quad \quad +\sum_{s=s^*+1}^{{\mathcal{S}}} \pi_s E[E\left\{u\left(A=1, \mathcal{T}_s, W\right) \mid W, A=1, S = s\right\} \\
&\quad \quad \quad \quad -E\left\{u\left(A=0, \mathcal{T}_s, W\right) \mid W, A=0, S = s\right\} \mid S = s ] 
\end{align*}

\end{document}